\documentclass{article}
\usepackage{spconf,amsmath,graphicx}
\usepackage{units}
\usepackage{balance}
\usepackage{units}
\usepackage{booktabs}
\usepackage{color}
\usepackage{amsmath,amssymb}

\def\L{{\mathcal{L}}}
\def\E{{\mathbb{E}}}

\graphicspath{{./Plots/}}

\title{ P\lowercase{ost}GAN: A GAN-B\lowercase{ased} P\lowercase{ost}-P\lowercase{rocessor to} E\lowercase{nhance the} Q\lowercase{uality of} C\lowercase{oded} S\lowercase{peech} }
\name{Srikanth Korse, Nicola Pia, Kishan Gupta, Guillaume Fuchs}
\address{
  Fraunhofer IIS, Erlangen, Germany\\ srikanth.korse@iis.fraunhofer.de\\
  }
  
\begin{document}

\ninept
\maketitle

\begin{sloppy} 
\begin{abstract}
  The quality of speech coded by transform coding is affected by various artefacts especially when bitrates to quantize the frequency components become too low. In order to mitigate these coding artefacts and enhance the quality of coded speech, a post-processor that relies on a-priori information transmitted from the encoder is traditionally employed at the decoder side. In recent years, several data-driven post-postprocessors have been proposed which were shown to outperform traditional approaches. In this paper, we propose PostGAN, a GAN-based neural post-processor that operates in the sub-band domain and relies on the U-Net architecture and a learned affine transform. It has been tested on the recently standardized low-complexity, low-delay bluetooth codec (LC3) for wideband speech at the lowest bitrate (16~\unit{kbit/s}). Subjective evaluations and objective scores show that the newly introduced post-processor surpasses previously published methods and can improve the quality of coded speech by around 20 MUSHRA points. 
\end{abstract}

\noindent\textbf{Index Terms}: Deep Neural Network (DNN), Speech Coding, Coded Speech Enhancement, Post-Filter, Post-Processor, Generative Adversarial Networks (GAN)

\section{Introduction} \label{sec:Introduction}
\paragraph*{}
The recently standardized low-complexity, low-delay codec (LC3)~\cite{LC3:2018Std, schnell2021lc3} relies on transform coding, quantizing and coding the spectral coefficients after a Modified Discrete Cosine Transform (MDCT). At medium to high bitrates, due to sufficient bits, transform coding yields sufficiently good to transparent quality. However, at low bitrates, many spectral coefficients are quantized to zero resulting in spectral holes, thereby causing audible artefacts commonly known as "birdie" artefacts~\cite{schnell2021lc3}. To enhance the perceptual quality at these low bitrates, tools such as noise filling~\cite{schnell2021lc3, disch2016intelligent} and Long Term Post-filter (LTPF) are employed~\cite{schnell2021lc3, Fuchs_2015, chen1995adaptive}. While noise filling typically aids in mitigating the audible artefacts by filling the spectral holes with pseudo-random noise scaled by a transmitted energy factor, the LTPF aims to improve the harmonicity of coded speech by attenuating inter-harmonic noise. LTPF depends on the pitch lag information provided by the encoder. This transmission of additional information from the encoder to the decoder as side information results in an overhead in the bit consumption. Moreover, these tools are tightly linked to the coding scheme and cannot be employed over an already deployed codec. 
  
Several data-driven models have been proposed as an alternative to enhance the quality of coded speech. While the majority of the models~\cite{Das2018, Zhao2019, Korse2020, Skoglund2019, Biswas2020} do not rely on any additional side information,~\cite{Hwang2021} transmits residual log power spectra to the decoder as side information. Among these data-driven models, generative models operating in time-domain are the most promising approach for recovering the lost spectral information during the quantization of the spectral coefficients to zero. However, this comes at the expense of complexity. Two prior works that consider generative models as post-processor, one~\cite{Skoglund2019} using an autoregressive model called LPCNet~\cite{valin19_interspeech}, and other named \textit{Deep Coded Audio Enhancer} (DCAE)~\cite{Biswas2020} opting for Generative Adversarial Networks (GANs). Since they were tested on two different codecs, direct comparison between autoregressive and GAN approaches in this context cannot be found in the literature.

However, in recent years, GANs have been shown to yield very high quality speech at very low computational cost, competing autoregressive models for applications like Text-to-Speech (TTS) and low bit rate speech coding~\cite{Kong_2020_HifiGAN, Mustafa2021_smgan, mustafa2021streamwise, Kumar2019MelGAN}. This serves as motivation to propose a GAN-based post-processor.

\subsection{Key Contributions of the Paper}

\begin{itemize}
\item We propose a GAN-based post-processor called \textit{PostGAN} that operates in the sub-band domain and combines the U-Net architecture with a learned affine transform to enhance the quality of coded speech.

\item We test our proposed model on the LC3 codec at 16~\unit{kbit/s} on wideband speech (16~\unit{kHz}) and is compared to the previously proposed methods in~\cite{Korse2020, Skoglund2019, Biswas2020}. 

\item PostGAN is suitable for commnunication applications, because, although trained on 1~\unit{s} frames during training, it can operate on 10~\unit{ms} frames during inference.

\item In contrast to~\cite{Skoglund2019} which requires access to the bitstream, our model, similar to~\cite{Korse2020, Biswas2020}, considered and processes only the coded speech at the output of the decoder. This allows our model to be deployed at the end of the communication chain with a minimal change.   

\item The proposed model is shown to outperform the prior methods for the given task. 

\end{itemize}

\section{Proposed Model}\label{sec:Model}
\subsection{Generator}

The generator architecture of the proposed PostGAN is shown in Figure~\ref{fig:Gen_Arch}. It takes coded speech $\tilde{x}$ and conditional features $h$ as input and outputs enhanced speech $\hat{x}$ by learning a mapping function $f(.)$ such that

\begin{equation}
\hat{x} =  f(\tilde{x}, h),
\label{equ:mapping_equ}
\end{equation}
 
\begin{figure}[t]
  \centering
  \includegraphics[width=\linewidth]{"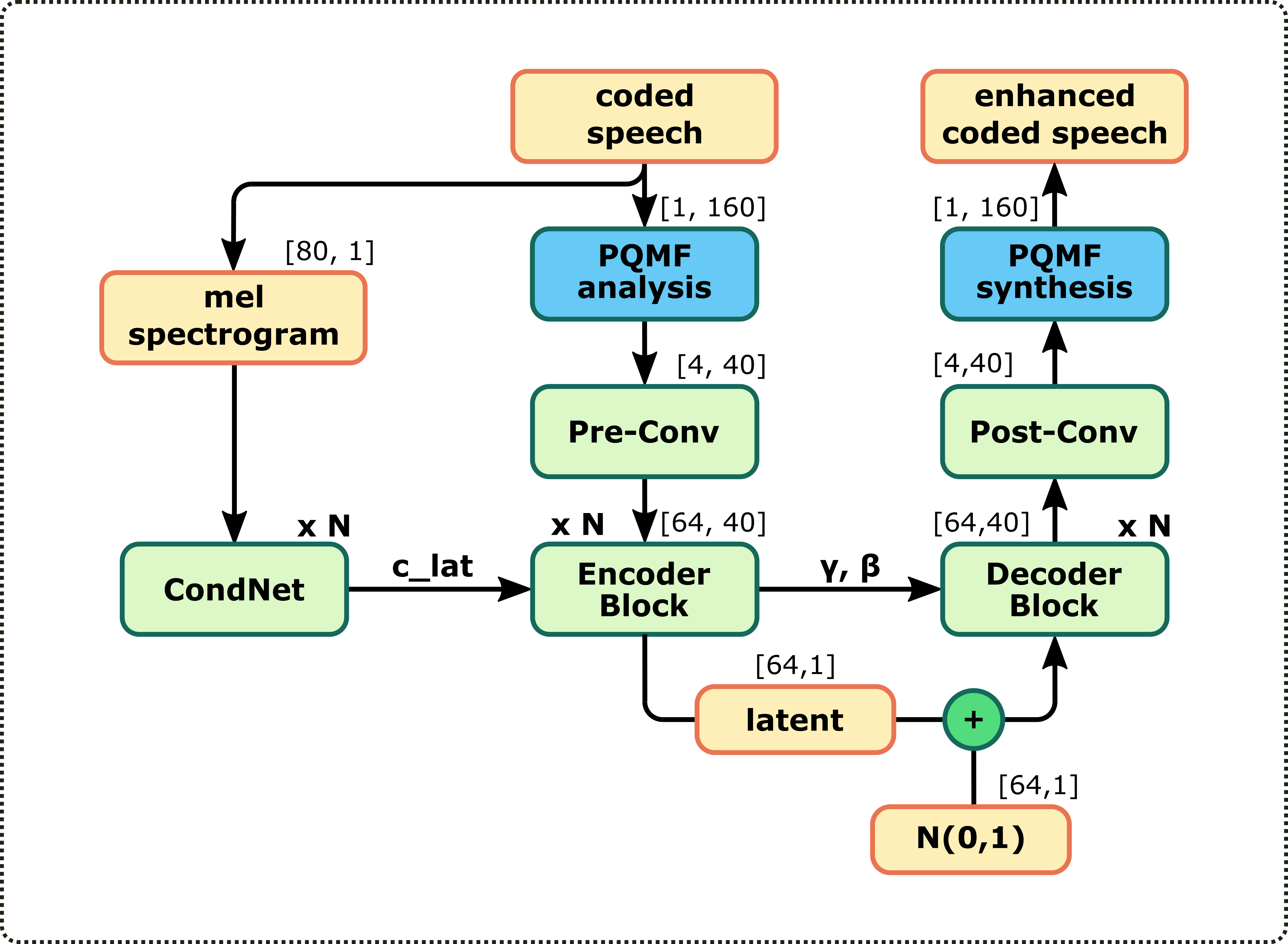"}
  \caption{Generator architecture of PostGAN with encoder, decoder and CondNet blocks along with PQMF analysis and synthesis. The value of N in the proposed architecture is 6.}
  \label{fig:Gen_Arch}
\end{figure}  

For the investigations in this paper, the conditional features $h$ comprise an 80-band mel-spectogram derived from the coded speech $\tilde{x}$.

To reduce computational complexity, the proposed generator operates in the subband domain similar to the ones in~\cite{Yang2021, mustafa2021streamwise}. The input time-domain signal is converted to subbands using PQMF analysis and the output subband signals are converted to time-domain signal using PQMF synthesis~\cite{Nguyen1994}. In addition, the generator network comprises of \textit{N} encoder, decoder and CondNet blocks along with pre-conv and post-conv layers. The detailed architecture of a single encoder, decoder, CondNet and up/downsample blocks are shown in Figures~\ref{fig:Gen_Enc_Dec_Blocks} and~\ref{fig:Gen_Arch_CondNet_UpDown}. In our implementation, we use 6 encoder, decoder and CondNet blocks and all the convolutions used are causal convolutions. 

\begin{figure}[t]
  \centering
  \includegraphics[width=\linewidth]{"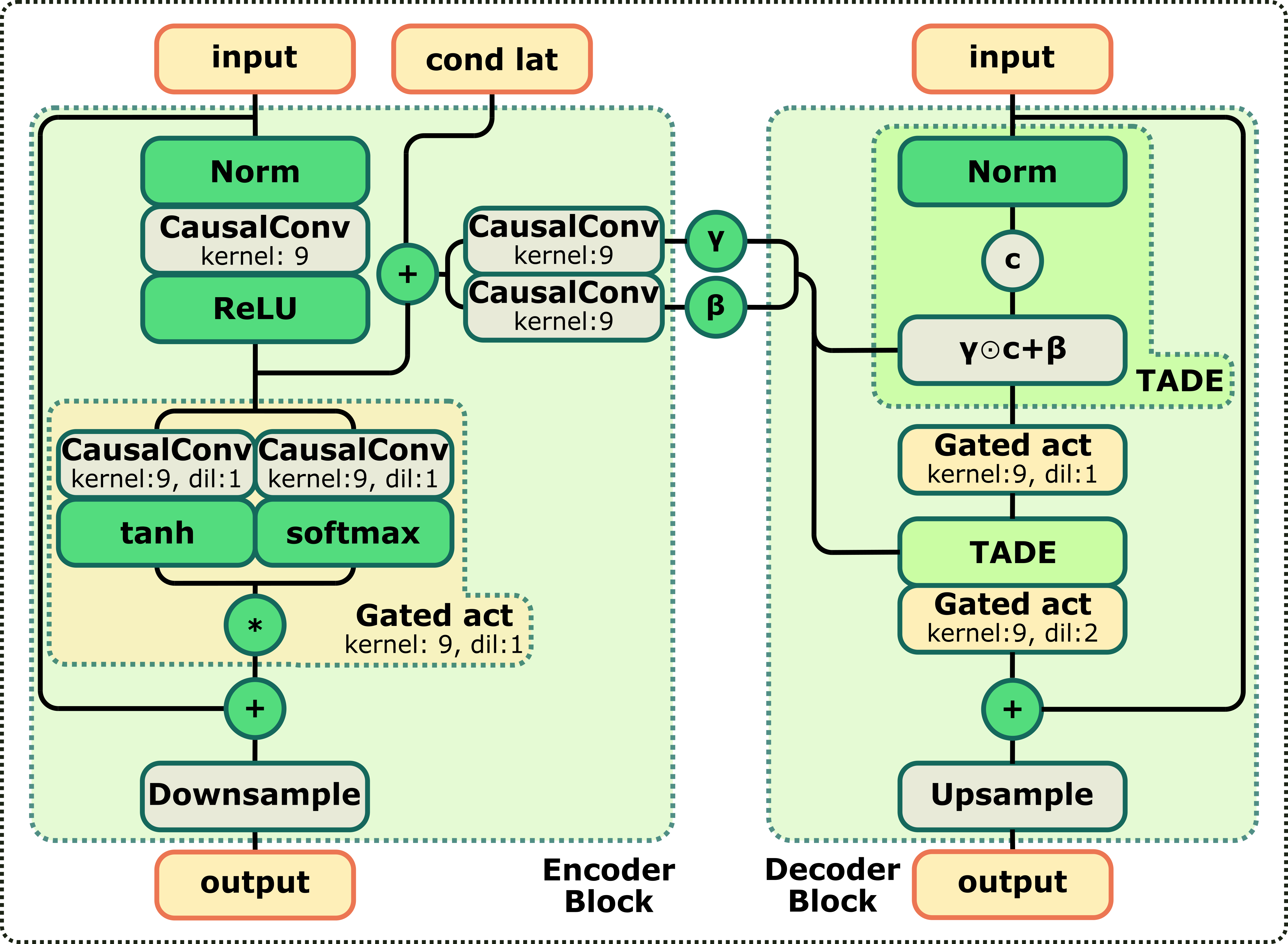"}
  \caption{Detailed description of encoder and decoder blocks.}
  \label{fig:Gen_Enc_Dec_Blocks}
\end{figure}

The main task of the encoder block is to compute the modulation parameters $\gamma$ and $\beta$ based on the internal latent representation of the encoder block and the latent obtained from the corresponding CondNet block. The information learnt by the model from conditional features and the input coded speech are complementary, i.e., the network learns the spectral envelope mainly from the conditional latent and finer temporal dependencies from the internal latent of the encoder block. This allows the network to mitigate artefacts such as pitch jumps and loudness mismatches that are common in parametric generative models. The internal latent representation is then passed through a gated activation followed by a downsampling block before passing it to the next encoder block. Since learning the parameters of an affine transform to modulate the latent was shown to outperform simple upsampling of latent~\cite{Mustafa2021_smgan, mustafa2021streamwise}, we use the same Temporal Adaptive DE-normalization (TADE) residual block as proposed in~\cite{Mustafa2021_smgan} as the basic building block of our decoder block. The only modification is that, the modulation parameters $\gamma$ and $\beta$ are not computed within the decoder block from the conditional features but obtained from the corresponding encoder block. All the normalization layers use channel normalization due to the benefits it brings compared to instance normalization as described in~\cite{mustafa2021streamwise}. The first decoder block takes as input the latent from last encoder block added with some random vector drawn from the normal distribution $\mathcal{N}(0,1)$. This was done to add some inherent stochastic part in the generator.  

\begin{figure}[t]
  \centering
  \includegraphics[width=\linewidth]{"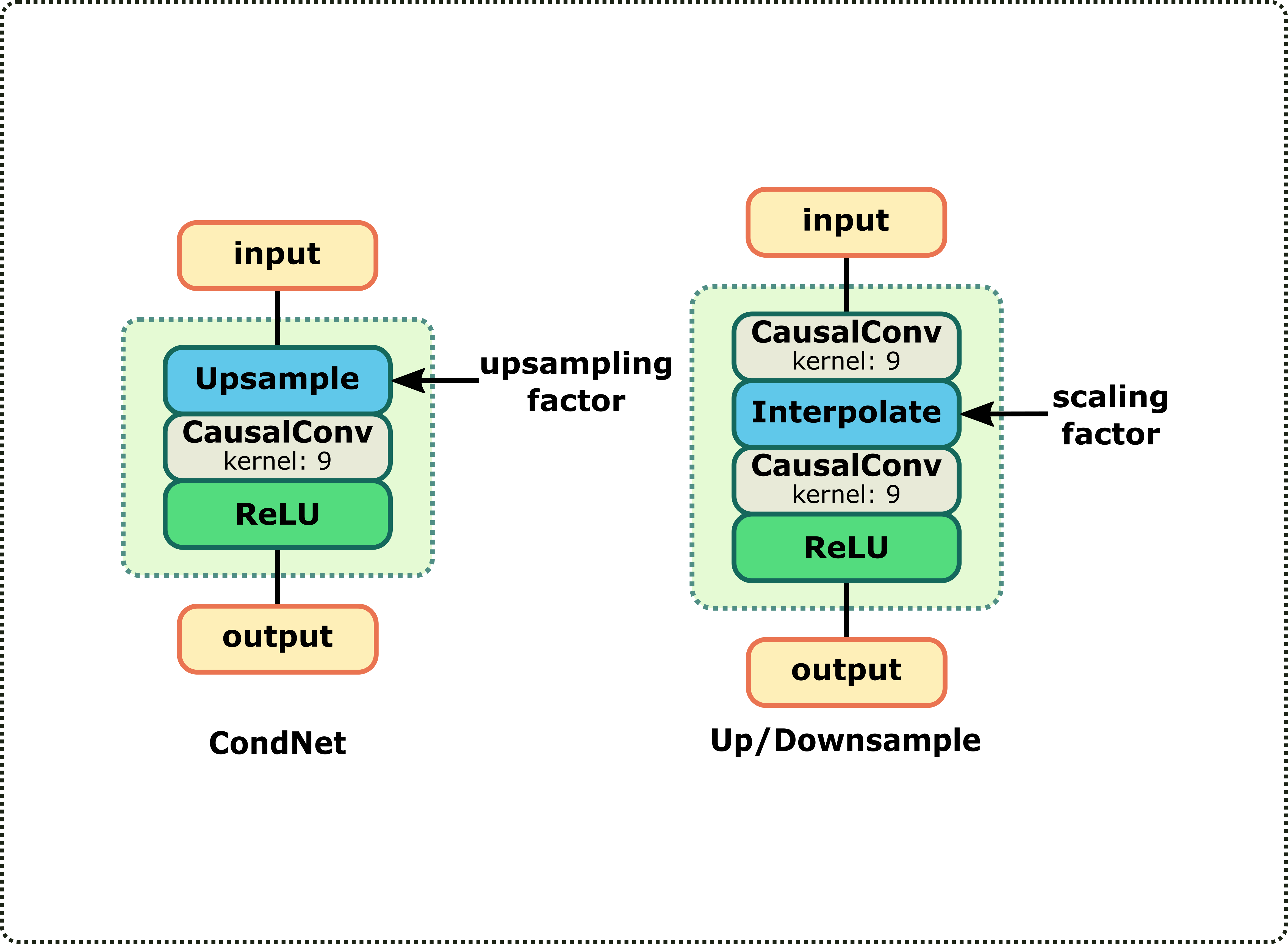"}
  \caption{Detailed description of CondNet and Up/Downsample blocks. The scaling factors used are [ 1, 2, 2, 2, 2.5, 2 ] and upsampling factors are [ 40, 40, 20, 10, 5, 2 ].}
  \label{fig:Gen_Arch_CondNet_UpDown}
\end{figure}

The CondNet learns hidden representations from the frame level conditional features by first upsampling followed by a causal convolutional layer and activation. Since each encoder block has a corresponding CondNet block, it allows the encoder to learn useful representations at different resolutions. The upsampling and downsampling blocks within the encoder/decoder blocks use an interpolation between two causal convolutional layers to either upsample or downsample the input. The main motivation to use such techniques instead of a simple transposed convolution was to avoid the artefacts caused by the upsampling~\cite{Pons_icassp_upsampling}. 

\subsection{Discriminator}
\label{subsec:disc}
For adversarial training, we use an ensemble of six discriminators operating on multiple random windowed slices of the input signal as proposed in~\cite{Kong_2020_HifiGAN}. The architecture of the individual discriminators is same as the one proposed in~\cite{Mustafa2021_smgan}. Three among six discriminators operate in the subband domain similar to~\cite{Mustafa2021_smgan} where a random window of length 512 is extracted from time-domain signal and is converted to 1, 2 and 4 bands, respectively, using PQMF analysis~\cite{Nguyen1994}. The other 3 discriminators are multi-scale discriminators~\cite{Kumar2019MelGAN} operating on the signal downsampled by a factor of 1, 2 and 4, respectively. The combination of subband and multi-scale discriminators was found to converge faster than either only subband or only multi-scale discriminators. 

\section{Experiments and Results}\label{sec:Expt_Setup}
\subsection{Training} \label{subsec:training}
The training procedure follows a two step strategy similar to the one used in~\cite{Mustafa2021_smgan, mustafa2021streamwise, Yamamoto2020} as it yields a stable and efficient training. At first, the generator is pre-trained using multi-resolution STFT loss $\L_{aux}$ between the log-magnitude of the output of the generator and the target signal at different STFT resolutions as described in Equation 6 of~\cite{Yamamoto2020}. The pre-training is followed by adversarial training where the adversarial metric comprises 2 components, hinge loss and the auxilary loss used as a regularization term. The final generative objective is given by

\begin{equation} \label{equ:gen_loss}
\min_{G} \left( \E_{\tilde{x}}\left[ \frac{1}{6} \sum_{k=1}^{6} -D_{k}(G(\tilde{x}, h)) \right]  + \L_{aux}(G) \right) , 
\end{equation}  

where $\tilde{x}$ is the coded speech and $h$ is the conditioning feature. All the convolutional layers of generator and $D_{k}$ use weight normalization~\cite{Saliman_WeightNorm_NIPS2016}

\subsection{Baseline Models}
For the evaluation, the proposed model PostGAN is compared to the following baseline models:

\begin{itemize}
\item An autoregressive generative model based on LPCNet~\cite{valin19_interspeech} as proposed in~\cite{Skoglund2019}. In contrast to the conditional features used in~\cite{Skoglund2019} we use the conditional features proposed in~\cite{valin19_interspeech} derived from the coded speech. Although suboptimal, this is done to keep the comparison fair. The training is also done in two steps as proposed in~\cite{Skoglund2019, valin19_interspeech}. Data augmentation is also turned off since it was observed that it did not bring any benefit on the database used for training. 

\item DCAE as proposed in~\cite{Biswas2020}. In order to avoid the mismatch of dimensions between the encoder and decoder of the U-Net architecture, we use the kernel size of 32 instead of 31 used in~\cite{Biswas2020}.

\item Improved DCAE which uses the same generator architecture but the discriminator used is the same as described in Section~\ref{subsec:disc}. The training of the model is done as described in Section~\ref{subsec:training}.

\item Mask-based post-filter~\cite{Korse2020} which estimates a real-valued mask per time-frequency bin. In contrast to~\cite{Korse2020} which operates only until 6.4~\unit{kHz} audio bandwidth, for our evaluation, we use the post-filter until 8~\unit{kHz}. 

\item Streamwise StyleMelGAN (SSMGAN) as proposed in~\cite{mustafa2021streamwise} but conditioned on a 80-band mel-spectogram derived from the coded speech. 

\end{itemize}

\subsection{Experimental Setup}
The PostGAN and the baseline models are trained on a NVIDIA Tesla V100 GPU on a subset of 30 speakers (15 male, 15 female) out of 109 speakers of VCTK corpus~\cite{Veaux_2013_VCTK} at 16~\unit{kHz}. LPCNet and DCAE apply a pre-emphasis filter on the input signal before feeding it to the model and then apply a de-emphasis on the enhanced signal. In contrast, the proposed PostGAN, SSMGAN and improved DCAE operate directly on the signal. The training methodology and hyperparameters of the mask-based model is same as~\cite{Korse2020}. For DCAE, the training hyperparameters are the same as in~\cite{Biswas2020} with the exception of batch size which is set to 32 and is trained for 150 epochs. The proposed PostGAN, SSMGAN and improved DCAE are pre-trained for 105k iterations followed by adversarial training for 645k iterations as explained in the Section~\ref{subsec:training} using the discriminator proposed in the Section~\ref{subsec:disc} on a batch size of 32. The generator is trained with the Adam optimizer with a learning rate of $lr_{g} = 1\cdot10^{-4}$ and $\beta=[0.5, 0.9]$ for the first 150 epochs. Then it is changed to $5\cdot10^{-5}$ but with $\beta$ unchanged. During the adversarial training, the discriminator is also trained with the Adam optimizer but with a learning rate of $lr_{d} = 5\cdot10^{-5}$ and same $\beta$.   

\subsection{Subjective Tests}

\begin{figure}[t]
  \centering
  \includegraphics[width=\linewidth]{"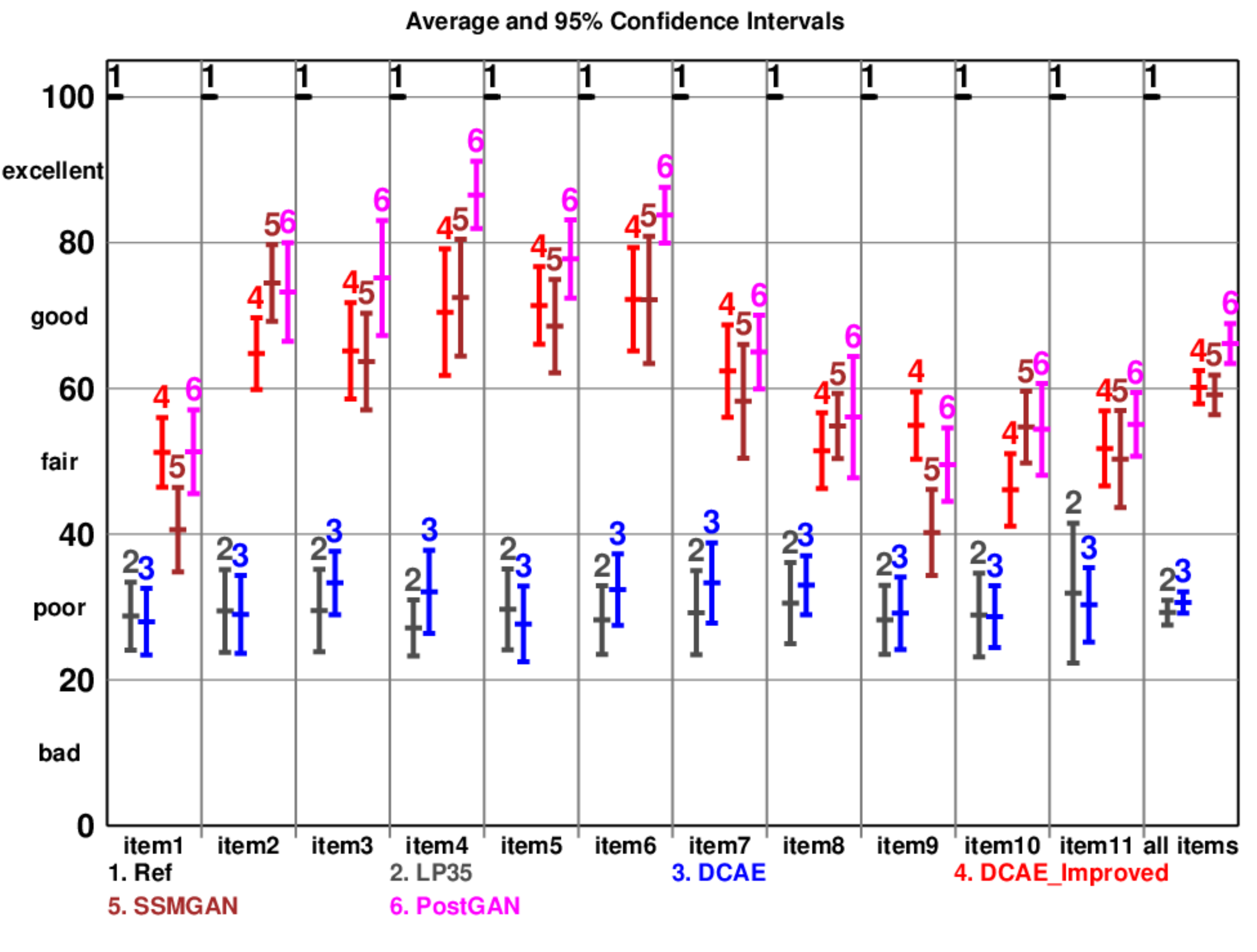"}
  \caption{Average MUSHRA scores with 13 listeners and Student's t distribution comparing PostGAN with other GAN-based models.}
  \label{fig:MUSHRA_1}
\end{figure}

For our subjective evaluation, we use the MUSHRA listening test methodology~\cite{MUSHRA}. The test is divided into two sub-tests as shown in Figures~\ref{fig:MUSHRA_1} and~\ref{fig:MUSHRA_2} with each test consisting of 13 listeners and 11 items. A 3.5kHz low-pass filtered version of the original signal was added as an anchor. Out of 11 items, 5 items come from the test set of the VCTK database and 6 items are from the unseen proprietary database consisting of 4 different languages (3 of the 4 languages are unseen during training). The first part of the test compares only the GAN-based models, i.e. PostGAN, SSMGAN, DCAE and improved DCAE. The results not only highlights the benefit of PostGAN over the other GAN-based models but also shows the disadvantage of the training method and the discriminator used to train DCAE in~\cite{Biswas2020}. The discriminator proposed in~\cite{Biswas2020} reaches saturation very early in the training, leading to the generator being trained only on the regularization term, i.e. L1 norm causing a low-pass effect. As the generators used in DCAE and improved DCAE are exactly the same, comparing them highlights the benefit of the discriminator proposed in Section~\ref{subsec:disc} and the training method proposed in Section~\ref{subsec:training}. Since the discriminator and the training method of improved DCAE and PostGAN are exactly the same, comparing them highlights the benefit of combining the U-Net architecture with an affine transform compared to a simple U-Net architecture. Comparing the SSMGAN with the PostGAN highlights the benefit of the additional information i.e. coded speech which helps in mitigating the artefacts caused by mismatch of pitch. 

The second part of the test compares the proposed PostGAN with LPCNet, Mask-based post-filter and LC3 (16~\unit{kbps}). From these results, we can conclude that: 1) LPCNet is suboptimal when operated just on the coded speech without access to the bitstream. 2) PostGAN successfully enhances the quality of coded speech and is better on average than the mask-based post-filter. However, like rest of the generative models, generalization to unseen speakers and languages is still problematic and is an area that needs to be addressed in future.

\begin{figure}[t]
  \centering
  \includegraphics[width=\linewidth]{"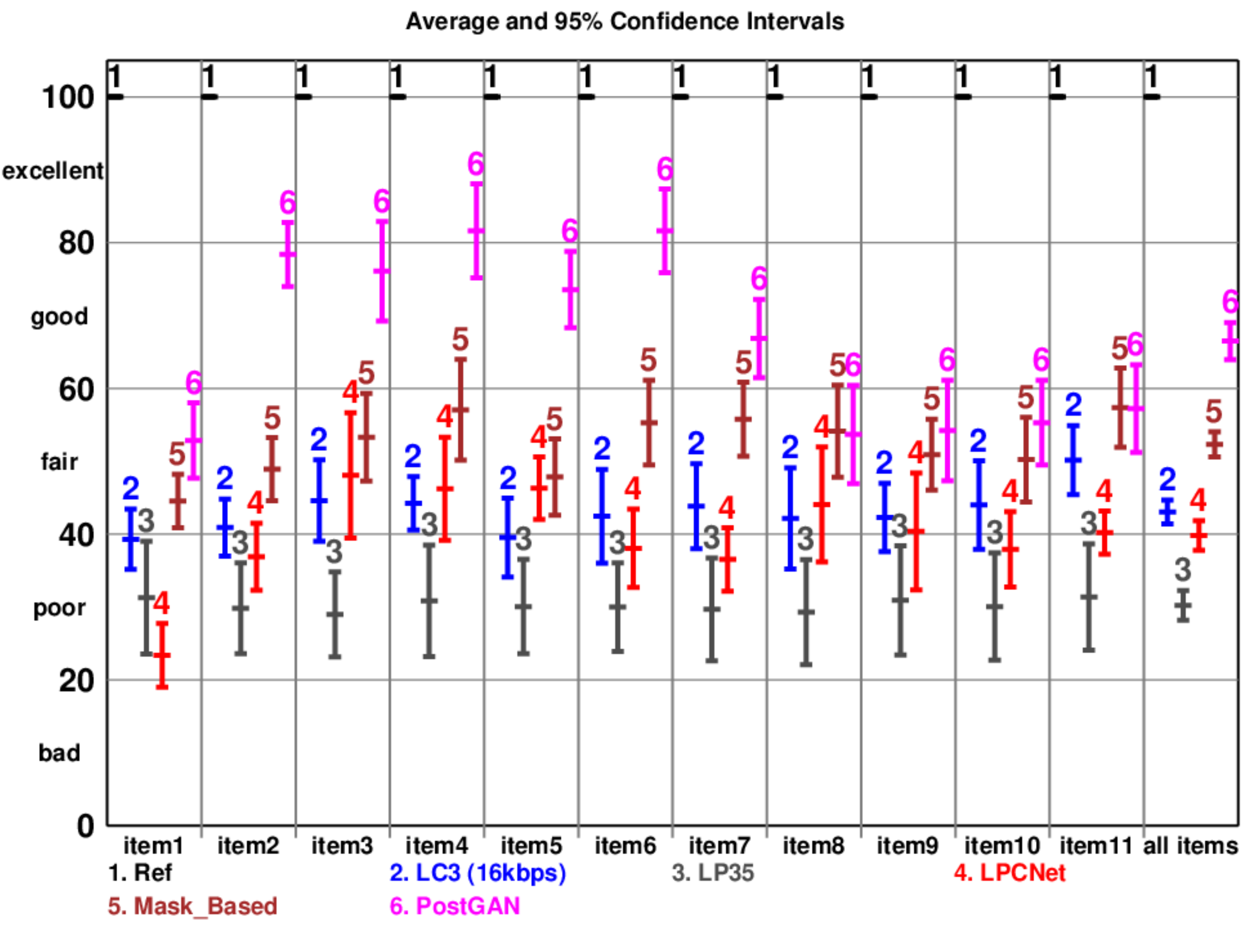"}
  \caption{Average MUSHRA scores with 13 listeners and Student's t distribution comparing PostGAN with prior methods.}
  \label{fig:MUSHRA_2}
\end{figure}

\subsection{Objective Tests}

\begin{table}[t]
\begin{center}
\begin{tabular}{  c  c  c  c  }
\hline
\textbf{Name} & \textbf{WARP-Q} & \textbf{STOI} & \textbf{POLQA}  \\ 
\hline
 LC3 & 0.6865  & 0.9280 & 3.4325 \\
 Mask-Based & \textbf{0.6244}  & 0.9529 & \textbf{4.081} \\
  LP35 & 0.6625 & \textbf{0.9999} & 3.9599 \\
 \hline
 PostGAN & \textbf{0.7107}  & \textbf{0.9258} & \textbf{3.7158} \\ 
 SSMGAN & 0.8041  & 0.8592 & 3.0993 \\
 DCAE & 0.7956  & 0.9242 & 3.4302 \\
 DCAE (Improved) & 0.8065  & 0.9177 & 3.4110 \\
 LPCNet & 0.8045  & 0.8219 & 2.9151 \\
 \hline

 \hline
\end{tabular}
 \caption{Average objective scores. Higher scores are better for STOI and POLQA and lower scores are better for WARP-Q. Confidence intervals are negligible }
 \label{tab:obj_scores} 
 \end{center}
\end{table}

\begin{table}[t]
\begin{center}
\begin{tabular}{  c  c  c  c   }
\hline
\textbf{Name} & \textbf{Complexity} & \textbf{Parameters}  &  \textbf{Delay} \\ 
  & (GMACs) & (Million) & (ms) \\
\hline
 Mask-Based & \textbf{0.8}  & \textbf{0.147} & 32  \\
 \hline
 PostGAN & 5.1  & 2.6  & \textbf{22.5} \\  
 SSMGAN & 4.8  & 2.77  & \textbf{22.5} \\
 LPCNet & 1.5  & 1.24 & 35 \\
 DCAE & 3.62 & 75.45 & 1024 \\
 \hline
\end{tabular}
  \caption{Table comparing PostGAN and baseline models in terms of Complexity, Number of Parameters and Delay.}
   \label{tab:complexity} 
  \end{center}
\end{table}

Objective measures such as POLQA~\cite{POLQA}, WARP-Q~\cite{Jassim2021_warpq} and STOI~\cite{Taal2011_STOI} are used to compare our proposed PostGAN with the baseline models. The obtained results are reported in Table~\ref{tab:obj_scores}. These scores were computed using a test set extracted from VCTK database comprising 2 speakers (1 male, 1 female) and 824 items. Since the mask-based post-filter is waveform preservering, the objective scores seem to prefer this compared to the generative models. Among the generative models, the proposed PostGAN, performs the best, whereas parametric models such as SSMGAN and LPCNet, which are non-waveform preserving perform the worst. In addition, all scores are also computed for the low-pass anchor (LP35). The scores for the low pass anchor clearly highlight that these objective scores are not completely reliable and do not fully reflect the perceived quality, especially for generative models which are non waveform preserving. 

\subsection{Complexity}

Table~\ref{tab:complexity} compares the proposed PostGAN with baseline models in terms of complexity, number of parameters and delay. Since mask-based model is a light-weight model, it has the lowest complexity and number of parameters compared to other models. However, it adds an additional delay of 32~\unit{ms} when operating in conjuntion to a low-delay codec such as LC3 which operates on a frame size of 10~\unit{ms}.

Among the generative models, although complexity of DCAE is comparable to other GAN based models, it is not possible to use it in a real-time system since it operates on a 1.024~\unit{sec} frames during inference. In addition, it requires almost 29x times more parameters compared to the proposed model. In contrast, the proposed PostGAN and SSMGAN operates on 10~\unit{ms} frames and adds an additional delay of 22.5~\unit{ms} due to the need of computation of mel-spectogram and PQMF analysis and synthesis allowing the models to be used in real-time communication. The delay of LPCNet is 35~\unit{ms} since it operates on 20~\unit{ms} frames with 50\% overlap and uses 2 lookahead frames. 

\section{Conclusion}\label{sec:Discussions_Conclusions}
We propose a GAN-based post-processor called PostGAN that is backward compatible to existing codecs and can operate in real-time during inference. It combines the U-Net architecture with a learned affine transformations, to directly enhance the coded speech in subband domain. For the investigations in the paper, we focused on mel-spectrogram computed from the coded speech as conditional features, although the model is flexible enough to be conditioned with features extracted from either the bitstream or from auxilliary information passed to the decoder. Our subjective tests and objective measure show that the proposed solution outperforms prior methods. In future, further complexity reduction and generalization will be addressed. 

\newpage

\bibliographystyle{IEEEtran}
\bibliography{refs19}

\begin{thebibliography}{10}
\providecommand{\url}[1]{#1}
\csname url@samestyle\endcsname
\providecommand{\newblock}{\relax}
\providecommand{\bibinfo}[2]{#2}
\providecommand{\BIBentrySTDinterwordspacing}{\spaceskip=0pt\relax}
\providecommand{\BIBentryALTinterwordstretchfactor}{4}
\providecommand{\BIBentryALTinterwordspacing}{\spaceskip=\fontdimen2\font plus
\BIBentryALTinterwordstretchfactor\fontdimen3\font minus
  \fontdimen4\font\relax}
\providecommand{\BIBforeignlanguage}[2]{{%
\expandafter\ifx\csname l@#1\endcsname\relax
\typeout{** WARNING: IEEEtran.bst: No hyphenation pattern has been}%
\typeout{** loaded for the language `#1'. Using the pattern for}%
\typeout{** the default language instead.}%
\else
\language=\csname l@#1\endcsname
\fi
#2}}
\providecommand{\BIBdecl}{\relax}
\BIBdecl

\bibitem{LC3:2018Std}
ESTI, ``{TR 103 590: Digital Enhanced Cordless Telecommunications (DECT); Study
  of Super Wideband Codec in DECT for narrowband, wideband and super-wideband
  audio communication including options of low delay audio connections},''
  {European Telecommunications Standards Institute (ETSI)}, TR {103 590}, 2018.

\bibitem{schnell2021lc3}
M.~Schnell, E.~Ravelli, J.~Büthe, M.~Schlegel, A.~Tomasek, A.~Tschekalinskij,
  J.~Svedberg, and M.~Sehlstedt, ``{LC3 and LC3plus}: The new audio
  transmission standards for wireless communication,'' in \emph{Audio
  Engineering Society Convention 150}, May 2021.

\bibitem{disch2016intelligent}
S.~Disch, A.~Niedermeier, C.~R. Helmrich, C.~Neukam, K.~Schmidt, R.~Geiger,
  J.~Lecomte, F.~Ghido, F.~Nagel, and B.~Edler, ``{Intelligent Gap Filling in
  Perceptual Transform Coding of Audio},'' in \emph{Audio Engineering Society
  Convention 141}, Sep 2016.

\bibitem{Fuchs_2015}
G.~{Fuchs}, C.~R. {Helmrich}, G.~{Marković}, M.~{Neusinger}, E.~{Ravelli}, and
  T.~{Moriya}, ``Low delay {LPC} and {MDCT}-based audio coding in the {EVS}
  codec,'' in \emph{2015 IEEE International Conference on Acoustics, Speech and
  Signal Processing (ICASSP)}, 2015, pp. 5723--5727.

\bibitem{chen1995adaptive}
{Juin-Hwey Chen} and A.~{Gersho}, ``Adaptive postfiltering for quality
  enhancement of coded speech,'' \emph{IEEE Transactions on Speech and Audio
  Processing}, vol.~3, no.~1, pp. 59--71, Jan 1995.

\bibitem{Das2018}
S.~Das and T.~B\"ackstr\"om, ``{Postfiltering Using Log-Magnitude Spectrum for
  Speech and Audio Coding},'' in \emph{Proc. Interspeech 2018}, 2018, pp.
  3543--3547.

\bibitem{Zhao2019}
Z.~{Zhao}, H.~{Liu}, and T.~{Fingscheidt}, ``{Convolutional Neural Networks to
  Enhance Coded Speech},'' \emph{IEEE/ACM Transactions on Audio, Speech, and
  Language Processing}, vol.~27, no.~4, pp. 663--678, April 2019.

\bibitem{Korse2020}
S.~Korse, K.~Gupta, and G.~Fuchs, ``{Enhancement of Coded Speech Using a
  Mask-Based Post-Filter},'' in \emph{ICASSP 2020 - 2020 IEEE International
  Conference on Acoustics, Speech and Signal Processing (ICASSP)}, 2020, pp.
  6764--6768.

\bibitem{Skoglund2019}
J.~Skoglund and J.-M. Valin, ``{Improving Opus Low Bit Rate Quality with Neural
  Speech Synthesis},'' in \emph{INTERSPEECH}, 2020.

\bibitem{Biswas2020}
A.~Biswas and D.~Jia, ``{Audio Codec Enhancement with Generative Adversarial
  Networks},'' in \emph{ICASSP 2020 - 2020 IEEE International Conference on
  Acoustics, Speech and Signal Processing (ICASSP)}, 2020, pp. 356--360.

\bibitem{Hwang2021}
S.~Hwang, Y.~Cheon, S.~Han, I.~Jang, and J.~W. Shin, ``{Enhancement of Coded
  Speech Using Neural Network-Based Side Information},'' \emph{IEEE Access},
  vol.~9, pp. 121\,532--121\,540, 2021.

\bibitem{valin19_interspeech}
J.-M. Valin and J.~Skoglund, ``{A Real-Time Wideband Neural Vocoder at 1.6kb/s
  Using LPCNet},'' in \emph{Proc. Interspeech 2019}, 2019, pp. 3406--3410.

\bibitem{Kong_2020_HifiGAN}
J.~Kong, J.~Kim, and J.~Bae, ``{HiFi-GAN: Generative Adversarial Networks for
  Efficient and High Fidelity Speech Synthesis},'' in \emph{NeurIPS}, 2020.

\bibitem{Mustafa2021_smgan}
A.~Mustafa, N.~Pia, and G.~Fuchs, ``{StyleMelGAN: An Efficient High-Fidelity
  Adversarial Vocoder with Temporal Adaptive Normalization},'' in \emph{ICASSP
  2021 - 2021 IEEE International Conference on Acoustics, Speech and Signal
  Processing (ICASSP)}, 2021, pp. 6034--6038.

\bibitem{mustafa2021streamwise}
A.~Mustafa, J.~Büthe, S.~Korse, K.~Gupta, G.~Fuchs, and N.~Pia, ``{A
  Streamwise GAN Vocoder for Wideband Speech Coding at Very Low Bit Rate},''
  \emph{IEEE Workshop on Applications of Signal Processing to Audio and
  Acoustics}, 2021.

\bibitem{Kumar2019MelGAN}
K.~Kumar, R.~Kumar, T.~Boissi{\`e}re, L.~Gestin, W.~Z. Teoh, J.~M.~R. Sotelo,
  A.~D. Br{\'e}bisson, Y.~Bengio, and A.~C. Courville, ``{MelGAN: Generative
  Adversarial Networks for Conditional Waveform Synthesis},'' in
  \emph{NeurIPS}, 2019.

\bibitem{Yang2021}
G.~Yang, S.~Yang, K.~Liu, P.~Fang, W.~Chen, and L.~Xie, ``{Multi-Band Melgan:
  Faster Waveform Generation For High-Quality Text-To-Speech},'' in \emph{2021
  IEEE Spoken Language Technology Workshop (SLT)}, 2021, pp. 492--498.

\bibitem{Nguyen1994}
T.~Nguyen, ``{Near-perfect-reconstruction pseudo-QMF banks},'' \emph{IEEE
  Transactions on Signal Processing}, vol.~42, no.~1, pp. 65--76, 1994.

\bibitem{Pons_icassp_upsampling}
J.~Pons, S.~Pascual, G.~Cengarle, and J.~Serrà, ``{Upsampling Artifacts in
  Neural Audio Synthesis},'' in \emph{ICASSP 2021 - 2021 IEEE International
  Conference on Acoustics, Speech and Signal Processing (ICASSP)}, 2021, pp.
  3005--3009.

\bibitem{Yamamoto2020}
R.~Yamamoto, E.~Song, and J.-M. Kim, ``{Parallel Wavegan: A Fast Waveform
  Generation Model Based on Generative Adversarial Networks with
  Multi-Resolution Spectrogram},'' in \emph{ICASSP 2020 - 2020 IEEE
  International Conference on Acoustics, Speech and Signal Processing
  (ICASSP)}, 2020, pp. 6199--6203.

\bibitem{Saliman_WeightNorm_NIPS2016}
T.~Salimans and D.~P. Kingma, ``{Weight normalization: A simple
  reparameterization to accelerate training of deep neural networks},'' in
  \emph{Proceedings of the 30th International Conference on Neural Information
  Processing Systems}, 2016.

\bibitem{Veaux_2013_VCTK}
C.~Veaux, J.~Yamagishi, and S.~King, ``{The voice bank corpus: Design,
  collection and data analysis of a large regional accent speech database},''
  in \emph{2013 International Conference Oriental COCOSDA held jointly with
  2013 Conference on Asian Spoken Language Research and Evaluation
  (O-COCOSDA/CASLRE)}, 2013, pp. 1--4.

\bibitem{MUSHRA}
{Recommendation BS.1534}, \emph{Method for the subjective assessment of
  intermediate quality levels of coding systems}, {ITU-R}, 2003.

\bibitem{POLQA}
\BIBentryALTinterwordspacing
\emph{Perceptual objective listening quality assessment ({POLQA})}, {ITU-T}
  Recommendation {P.863}, 2011. [Online]. Available:
  \url{http://www.itu.int/rec/T-REC-P.863/en}
\BIBentrySTDinterwordspacing

\bibitem{Jassim2021_warpq}
W.~A. Jassim, J.~Skoglund, M.~Chinen, and A.~Hines, ``{Warp-Q: Quality
  Prediction for Generative Neural Speech Codecs},'' in \emph{ICASSP 2021 -
  2021 IEEE International Conference on Acoustics, Speech and Signal Processing
  (ICASSP)}, 2021, pp. 401--405.

\bibitem{Taal2011_STOI}
C.~H. Taal, R.~C. Hendriks, R.~Heusdens, and J.~Jensen, ``{An Algorithm for
  Intelligibility Prediction of Time–Frequency Weighted Noisy Speech},''
  \emph{IEEE Transactions on Audio, Speech, and Language Processing}, vol.~19,
  no.~7, pp. 2125--2136, 2011.

\end{thebibliography}

\end{sloppy}
\end{document}